%% file: topq.tex
\documentclass[aps,prl,twocolumn,superscriptaddress,showpacs,floatfix]{revtex4}
\usepackage{graphicx}
\usepackage{dcolumn}
\usepackage{bm}
\usepackage{lineno}
\usepackage{setspace}
\usepackage{subfig}
\usepackage{booktabs}
\usepackage{rotating}
\usepackage{amssymb}

\newcommand\et{E_T}
\newcommand\pt{p_T}
\newcommand\Ht{H_T}
\newcommand\ttbar{t\bar{t}}
\newcommand\ppbar{p\bar{p}}
\newcommand\bbbar{b\bar{b}}

\begin{document}


\title{Exclusion of an Exotic Top Quark with $-$4/3 Electric Charge Using Soft Lepton Tagging}
\input{October2009_Authors.tex}
%

%

\begin{abstract}
We present a measurement of the electric charge of the top quark using $\ppbar$ collisions corresponding to an integrated luminosity of 2.7~fb$^{-1}$ at the CDF II detector.  We reconstruct $\ttbar$ events in the lepton+jets final state and use kinematic information to determine which $b$-jet is associated with the leptonically- or hadronically-decaying $t$-quark.   Soft lepton taggers are used to determine the $b$-jet flavor.  Along with the charge of the $W$ boson decay lepton, this information permits the reconstruction of the top quark's electric charge.  Out of 45 reconstructed events with $2.4\pm0.8$ expected background events, 29 are reconstructed as $\ttbar$ with the standard model $+$2/3 charge, whereas 16 are reconstructed as $\ttbar$ with an exotic $-4/3$ charge.  This is consistent with the standard model and excludes the exotic scenario at 95\% confidence level.  This is the strongest exclusion of the exotic charge scenario and the first to use soft leptons for this purpose.
\end{abstract}

\maketitle
%

%

Since the discovery of the top quark in 1995~\cite{TopDiscovery}, the CDF and D0 collaborations have scrutinized its properties.  Measurements of the properties of the top quark all present a consistent picture of the top quark as the third-generation standard model (SM) weak-isospin partner of the bottom quark \cite{TopProp}.  However a $+$2/3-electric-charged top quark has yet to be experimentally confirmed, and an exotic $-$4/3-charged scenario has been proposed~\cite{XMTheory}. In this theoretical scenario, the observed excess of events historically attributed to the top quark are instead attributed to an exotic particle, called ``XM top quark,'' which is identical to the SM top quark except that it decays to $W^- b$ rather than to the SM $W^+ b$.  In order to preserve anomaly cancellation, the weak-isospin partner of the bottom (i.e. the ``true'' top quark) is assumed to exist but is too massive to be observed experimentally.

In this Letter, we present a measurement of the electric charge of the top quark.  We analyze data corresponding to an integrated luminosity, $\int\mathcal{L}dt=2.7$~fb$^{-1}$, collected from February 2002 to April 2008.

We measure the top-quark charge by reconstructing $\ttbar$ pairs in the $\ell\bar{\nu} \bar{b}\, q\bar{q}'b$ final state.  The $b$-quarks associated with the leptonically (hadronically) decaying $W$ are called the ``leptonic'' (``hadronic'') $b$-quarks.  Reconstructing the top-quark charge involves identifying either the leptonic or hadronic $b$-quark and determining its flavor, either as $b$ or $\bar{b}$.  We use a soft electron tagger (SLT$_e$) \cite{SLTe} and a soft muon tagger (SLT$_\mu$) \cite{SLTmu}  (collectively referred to as the SLT taggers) to identify the $b$-jets.  The charges of the soft leptons are used to infer the flavor of the $b$-jets.  A secondary vertex tagger (\textsc{SecVtx}) \cite{Secvtx} is also used to identify the $b$-jets and suppress SM backgrounds.  A kinematic fitter \cite{KinFit} determines which $b$-jet is leptonic and which is hadronic.  An event is considered SM if the lepton from the $W$ and the SLT lepton from the leptonic (hadronic) $b$-jet have the opposite (same) charge sign.  The event is considered to be XM otherwise.  For the purposes of this measurement, we assume that the XM top quark has identical properties to the SM top quark, except for its electric charge.

This binary event reconstruction implies that if both the kinematic fitter and the SLT tagger are incorrect, then the correct top-quark charge is still reconstructed.  From a Monte Carlo (MC) simulation of $\ttbar$ events, the fraction of $b$-jets for which the SLT taggers give the correct flavor assignment is approximately $69\%$.  The fraction of events for which the kinematic fitter properly determines whether a $b$-jet is the leptonic or hadronic $b$ is approximately $76\%$.  This method reconstructs a SM (XM) charge in approximately 60\% (40\%) of simulated SM $\ttbar$ events.

This technique complements the measurement of the top-quark charge in Ref.~\cite{JetQ} which uses the curvature and momentum of tracks within a $b$-jet cone to determine its charge.  The SLT method is much less efficient than this technique since the semileptonic branching fraction for $b$-jets is only $\sim\!10\%$ per lepton flavor; however, the $b$-jet flavor determination is much more reliable because of the higher $b$-jet flavor reconstruction purity.  The overall reduction in sensitivity with the SLT technique is therefore only a factor of $2-3$ lower.

The CDF II detector \cite{CDF} is an azimuthally and forward-backward symmetric general-purpose detector~\cite{Eta} with silicon tracking~\cite{SVX} and drift chamber tracking~\cite{COT} immersed in a 1.4~T solenoidal magnetic field.  Projective electromagnetic- and hadronic-sampling calorimeters \cite{Calo} lie beyond the solenoid and provide jet and missing $\et$ ($\displaystyle{\not}{\et}$) reconstruction.  Muon chambers~\cite{Muons} lie beyond the calorimeter and provide coverage up to $|\eta|\le 1.0$.  Gaseous Cherenkov counters \cite{CLC} measure the average number of inelastic $\ppbar$ collisions to determine the luminosity with a 6\% relative uncertainty.

Events are identified with central  $(|\eta|\!\lesssim1)$, high-$\pt$ (-$\et$) muon (electron) triggers.  We select events with a $\pt>20$~GeV/$c$ ($\et>20$~GeV) muon (electron), which we call the ``primary'' lepton.  At least four jets \cite{JetReco} with corrected $\et>20$~GeV~\cite{JES} and $|\eta|\le2.0$ must be present in the event.  To increase our acceptance for $\ttbar$ events, we allow one of the four jets to pass a looser selection ($\et>12$~GeV and $|\eta|\le2.4$), but we do not consider the looser fourth jet for tagging, either by the SLT or \textsc{SecVtx} algorithms.  We explicitly reject cosmic muons, electrons from photon conversions, leptons from $Z$ boson decay, and events with more than one energetic and isolated lepton.  We also require $\Ht>250$~GeV and $\displaystyle{\not}{\et}>30$~GeV, where $\Ht$ is the scalar sum of the transverse energy of the primary lepton, $\displaystyle{\not}{\et}$, and jets.

We require each event to have $\ge 1$ SLT (either e or $\mu$) tag, and $\ge 1$ \textsc{SecVtx} tag.  We do not require that different jets in the same event are tagged by the different taggers.  To suppress cascade decays of $b$-jets (i.e. $b\rightarrow c \rightarrow\ell\nu X$) that result in flavor mis-identification, we require the SLT track $\pt>6$~GeV/$c$, since leptons from cascade decays tend to be softer than those from direct semileptonic decays.  We further require $p^{\rm rel}_T > 1.5$~GeV/$c$ where $p^{\rm rel}_T$ is the SLT$_\mu$ track $\pt$ relative to the jet axis.

We use a kinematic fitter described in detail in Ref.~\cite{KinFit} which minimizes a reduced $\chi^2$-like function to fit to the $\ttbar$ event hypothesis.  The experimental resolution of the final state particles is accounted for, and the particles are kinematically constrained to the $W$ mass and top-quark mass (assumed to be 175~GeV/$c^2$), within the theoretical decay widths.  Jets are assigned uniquely to each of the four final-state quarks, and those jets tagged by either the SLT or \textsc{SecVtx} algorithms are constrained to be either of the two $b$-jets.  All possible permutations are considered and the one which results in the lowest $\chi^2$ value is chosen.  If two different jets are both tagged, then we require that the lowest $\chi^2<27$; however if only one jet in the event is tagged, by both \textsc{SecVtx} and the SLT, then we require $\chi^2<9$.  The tighter requirement on the $\chi^2$ enforces a higher top-quark charge reconstruction purity since there is a greater ambiguity when only one jet is identified as a $b$-jet by the taggers.

The requirement on the $\chi^2$, SLT track $\pt$, and SLT$_\mu$ $p^{\rm rel}_T$ variables is determined by optimizing on total expected $\epsilon D^2$, where $\epsilon$ is the event-reconstruction efficiency, $D=2P-1$ is the dilution, where $P$ is the purity, which is defined as the fraction of reconstructed events that are determined to have an SM charge.  Table~\ref{table:merit} presents the expected $\epsilon D^2$ using the \textsc{pythia} MC generator~\cite{Pythia} to model $\ttbar$ and assuming $\sigma_{t\bar{t}}=6.7\pm0.8$~pb~\cite{TopXSTheory}, $M_{t}=175$~GeV/$c^2$, and $\int{\mathcal{L}}dt=2.7$~fb$^{-1}$.  We choose the pretag expectation as the denominator of the efficiency, although for the optimization that choice is arbitrary.  The figure of merit, $\epsilon D^2$, is shown for the combined result, as well as separately for events with one or two tagged jets and with the SLT$_e$ or SLT$_\mu$ only.  We expect $30.0\pm 5.9$ events from $\ttbar$ in the tag sample, where the uncertainty is dominated by the theoretical cross section uncertainty and the jet energy scale uncertainty.

\begin{table}[ht]
\centering
\begin{tabular}{cccccc}
\hline
\hline
 & $\epsilon$ (\%) & $P$ (\%) & $\epsilon D^2$ (\%) & $\langle N_{SM}\rangle$ & $\langle N_{XM}\rangle$ \\
\hline
 Total & 3.26 & 60.8 & 0.152 & 18.3 & 11.8 \\
\hline
1 tagged jet & 0.92 & 58.2 & 0.025 & 4.9 & 3.5 \\
$\ge 2$~tagged jets & 2.34 & 61.8 & 0.130 & 13.4 & 8.3 \\
\hline
SLT$_\mathrm{e}$ only & 1.62 & 61.9 & 0.092 & 9.2 & 5.7 \\
SLT$_\mu$ only & 1.69 & 59.4 & 0.060 & 9.3 & 6.3 \\
\hline
\hline
\end{tabular}
\caption{Expected efficiency ($\epsilon$), purity ($P$), $\epsilon D^2$, and number of events reconstructed as SM and XM, assuming $\sigma_{t\bar{t}}=6.7$~pb for $\int{\mathcal{L}}dt=2.7$~fb$^{-1}$ in \textsc{pythia} simulation.}
\label{table:merit}
\end{table}

Due to the requirement of at least two $b$-tags, the contribution from non-$\ttbar$ backgrounds to the data is very small.  Backgrounds from $W$+jets, $WW$, $WZ$, $ZZ$, single top, $Z$+jets, Drell-Yan+jets, and multijet production are all considered.  The dominant background is due to $W+\bbbar$ production, and the total expected contribution to the tag sample from all backgrounds is $2.4\pm0.8$ events, where the uncertainty is dominated by the uncertainty on the jet energy scale and the multijet background estimate.  The background estimate uses the same technique as in Ref.~\cite{SLTe}.

The measurement of the $b$-jet flavor determination purity is estimated with simulation but is calibrated by comparing a sample of pure $b\bar{b}$ events in both data and MC.  The sample is constructed from events with a $\pt>8$~GeV$/c$ ($\et>8$~GeV) muon (electron) close to a jet.  A recoiling jet must also be found in the event in which an SLT tag is present, and both jets must be tagged by \textsc{SecVtx}.  We measure a dilution scale factor, $SF_D=\sqrt{D_{\rm data}/D_{\rm MC}}$ where $D=(\epsilon_{\rm OS}-\epsilon_{\rm SS})/(\epsilon_{\rm OS}+\epsilon_{\rm SS})$, and $\epsilon_{\rm OS/SS}$ is the tagging efficiency when the trigger lepton and the SLT have the opposite sign (OS) or same sign (SS) charge.  The square root originates from the assumption that since both $b$-jets decay semileptonically they are subject to the same dilution factor.  This scale factor accounts for differences between the data and MC such as mis-estimated branching ratios or mis-modeling of neutral $B$ mixing.

We measure $SF_D$ to be $0.92\pm0.11$.  The uncertainty is dominated by the statistical uncertainty and covers dependencies on other variables, such as the jet $\et$.  We use this to correct the simulation estimate for the $\ttbar$ charge reconstruction purity, for which our final estimate is $(60\pm3)\%$.  The uncertainty in the purity is dominated by uncertainties in the simulation of the QCD radiation both in the initial (ISR) and final-state (FSR), the uncertainty in $SF_D$, and that arising when an alternate MC generator, \textsc{herwig} \cite{Herwig}, is used.  Simulation confirms the naive expectation that the background is reconstructed symmetrically between SM and XM events.  Although simulation may mis-model the background reconstruction, the total background contribution is small, so we apply a very conservative systematic uncertainty, corresponding to twice the uncertainty on the signal purity estimate.  Therefore, the background has a ``purity'' of $(50\pm6)\%$.

\begin{table}[ht]
\centering
\begin{tabular}{ccccc}
\hline
\hline
Subsample & $N$ & $N_{SM}$ & $N_{XM}$ \\
\hline
Primary Electron & 25 & 16 & 9 \\
Primary Muon & 20 & 13 & 7 \\
\hline
1 Tagged Jet & 7 & 4 & 3 \\
$\ge 2$~Tagged Jets & 38 & 25 & 13 \\
\hline
SLT$_e$ & 25 & 15 & 10 \\
SLT$_\mu$ & 21 & 15 & 6 \\
\hline
All & 45 & 29 & 16 \\
\hline
\hline
\end{tabular}
\caption{Tag configurations in various subsamples of the data, including divisions according to the primary lepton flavor, the number of tagged $b$-jets, and the SLT flavor. Shown are the number of SM and XM tags as well as the total.}
\label{table:asym}
\end{table}

We observe 45 tagged and reconstructed events in data, of which 29 are reconstructed as SM and 16 are reconstructed as XM, a ratio consistent with the SM hypothesis.  Three events have two SLT tags, although only one of these events has both SLT tags close to jets identified as $b$-jets by the kinematic fit (in this case, both SLT tags are consistent with the SM).  Table~\ref{table:asym} shows the number of tags by subsample, including the flavor of the primary lepton, the number of tagged $b$-jets, and the SLT flavor.  Note that there is no significantly different SM/XM admixture in any of the subsamples.

The statistical significance of the measurement is given by the $p$-value for the test statistic
\begin{equation}
A\equiv\frac{1}{D_S}\frac{N_{SM}-N_{XM}-\langle B\rangle D_B}{N_{SM}+N_{XM}-\langle B\rangle}
\end{equation}
where $N_{SM}$ ($N_{XM}$) is the number of SM (XM) events, $D_S$ and $D_B$ are the signal and background dilution, respectively, and $\langle B\rangle$ is the total background expectation.  This asymmetry, $A$, has been normalized so that the median expectation of the  SM (XM) hypothesis is $+1.0$ ($-1.0$).  In the data, we measure a normalized asymmetry $A_0=1.53\pm0.75$\,(stat), which clearly favors a SM hypothesis.

\begin{figure}[t]
\centering
\includegraphics[width=2.5in]{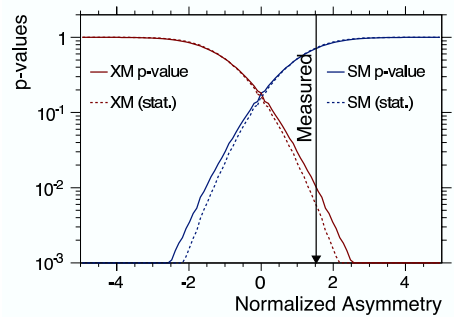}
\caption{The SM and XM $p$-values for the normalized asymmetry test statistic, $A$, from pseudo experiments shown with all uncertainties combined and statistical uncertainties only.}
\label{fig:pvalues}
\end{figure}

We use MC pseudo experiments to determine the $p$-value.  This is done by drawing from Poisson and binomial distributions to model the expected number of events and purities, respectively.  Uncertainties are treated as Gaussian distributions.  We measure $p_{\rm SM}=p(A\leq A_{0} | {\rm SM})=0.69$ and $p_{\rm XM}=p(A\geq A_{0} | {\rm XM})=0.0094$ for the SM and XM hypotheses, respectively, while we expect $p_{\rm SM}=0.50$ and $p_{\rm XM}=0.028$, assuming the SM.  Figure~\ref{fig:pvalues} shows the distribution of $p$-values under the SM and XM hypotheses from pseudo-experiments.  We choose the type-I error rate, $\alpha$, \textit{a priori} by using the standard threshold for exclusion of exotica: $\alpha=0.05$.  From this we exclude the exotic $-$4/3-charged top quark at 95\% confidence level.  Table~\ref{table:syst} shows the expected and measured XM $p$-value with the significant systematic errors added cumulatively.

\begin{table}[!ht]
\centering
\begin{tabular}{ccccc}
\hline\hline
Source  & Expected $p$-value & Observed $p$-value \\
\hline
Stat. only & 0.020 & 0.0054 \\
Dilultion Scale Factor & 0.021 & 0.0058 \\
ISR/FSR & 0.022 & 0.0062 \\
Cross Sections & 0.023 & 0.0069 \\
Jet Energy Scale & 0.026 & 0.0080 \\
MC Generator & 0.028 & 0.0094 \\
\hline\hline
\end{tabular}
\caption{The $p$-values obtained for the XM hypothesis and how they are affected by the cumulative addition of systematic uncertainties.  Other sources of systematic uncertainties are negligible.}
\label{table:syst}
\end{table}

We can also quantify the result of this measurement with a Bayes Factor (BF), which can be interpreted as the posterior odds in favor of the SM when the prior odds are neutral (equal to unity).  This quantity is equal to the ratio, $p(A=A_0|{\rm SM})/p(A=A_0|{\rm XM})$.  We evaluate a BF for this measurement to be 85.8 which is considered ``strong'' evidence  \cite{BF} for a $+$2/3-charged top quark.

\begin{figure}[b]
\centering
\includegraphics[width=1.5in]{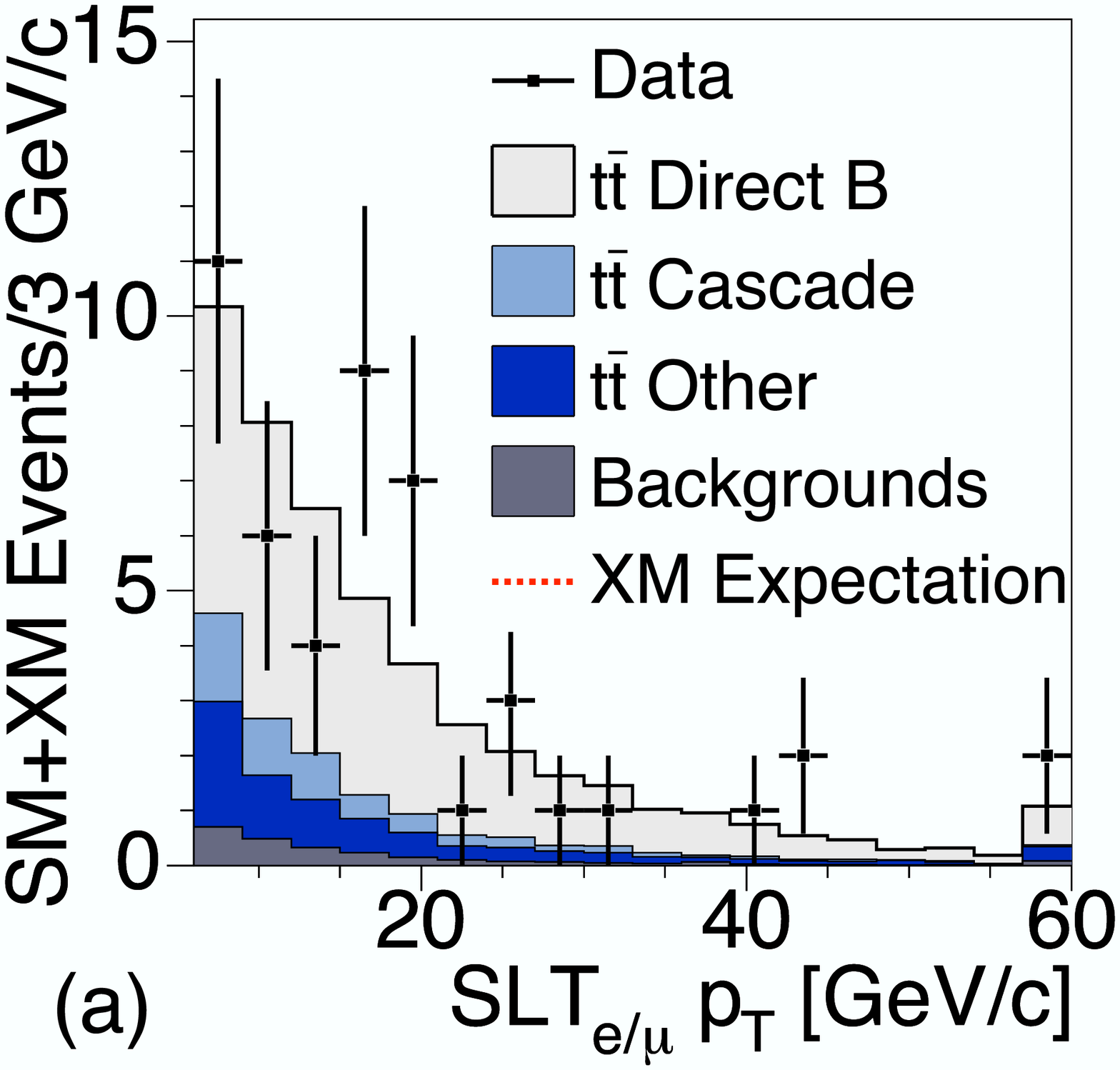}
\includegraphics[width=1.5in]{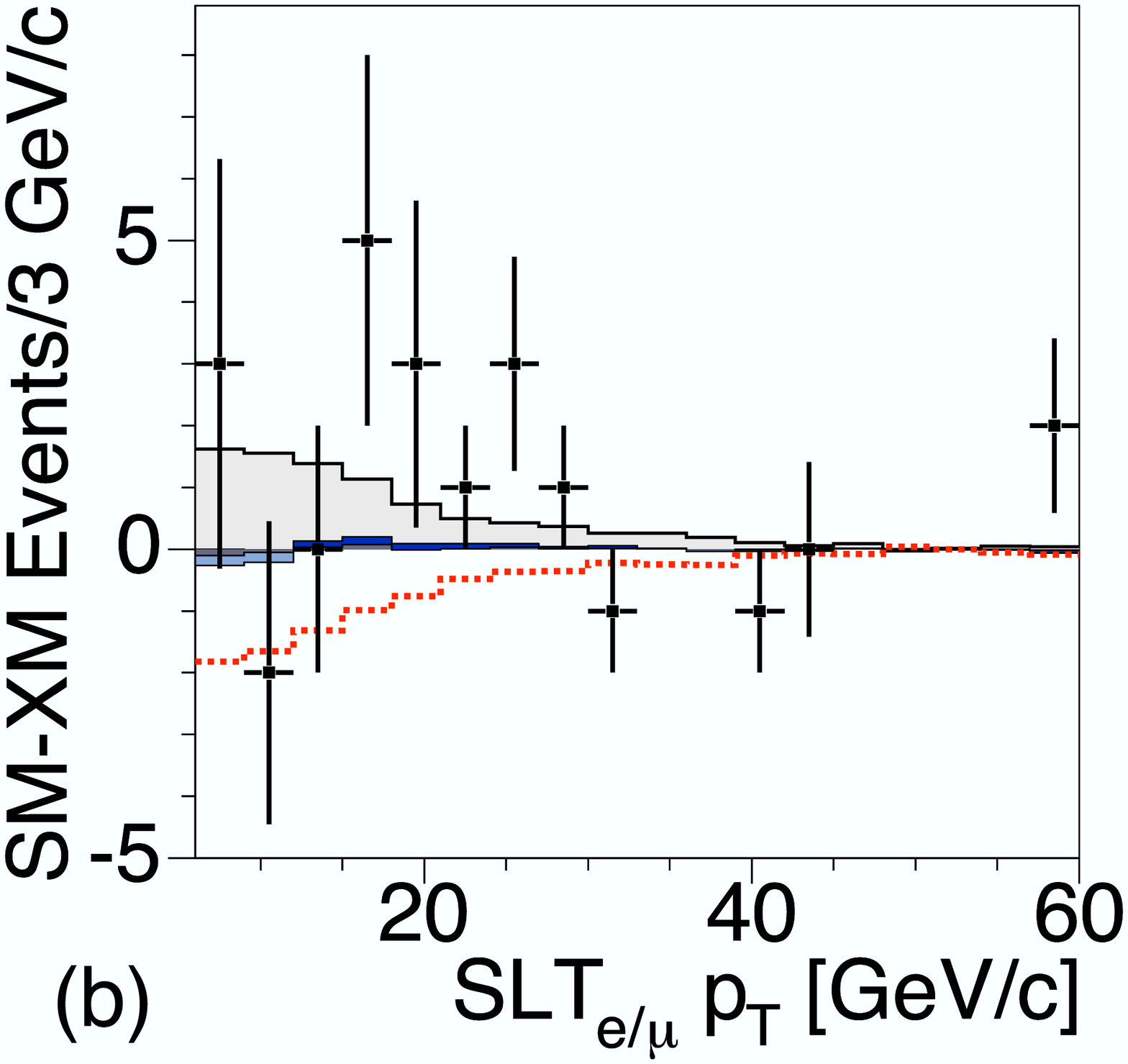}
\includegraphics[width=1.5in]{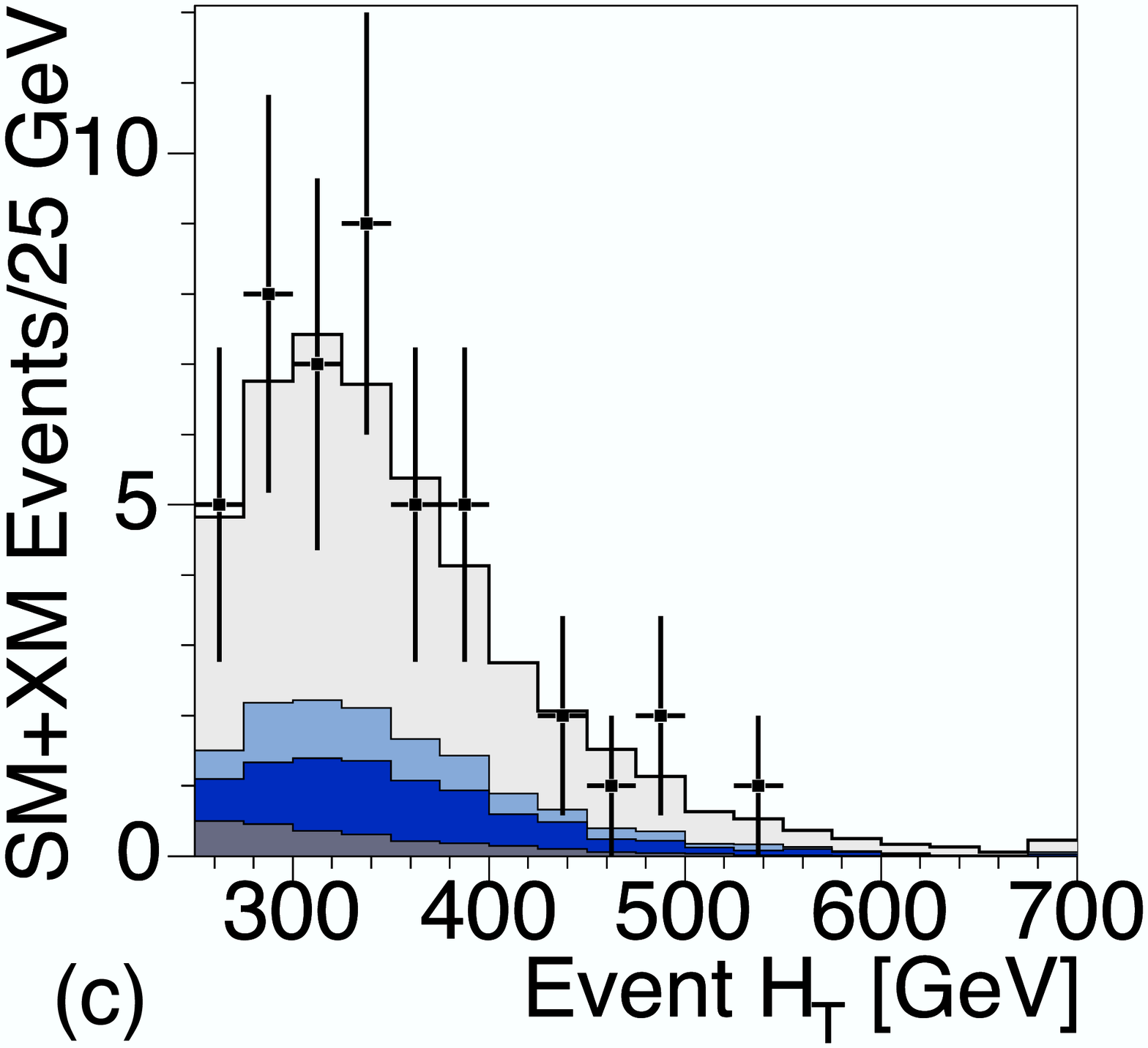}
\includegraphics[width=1.5in]{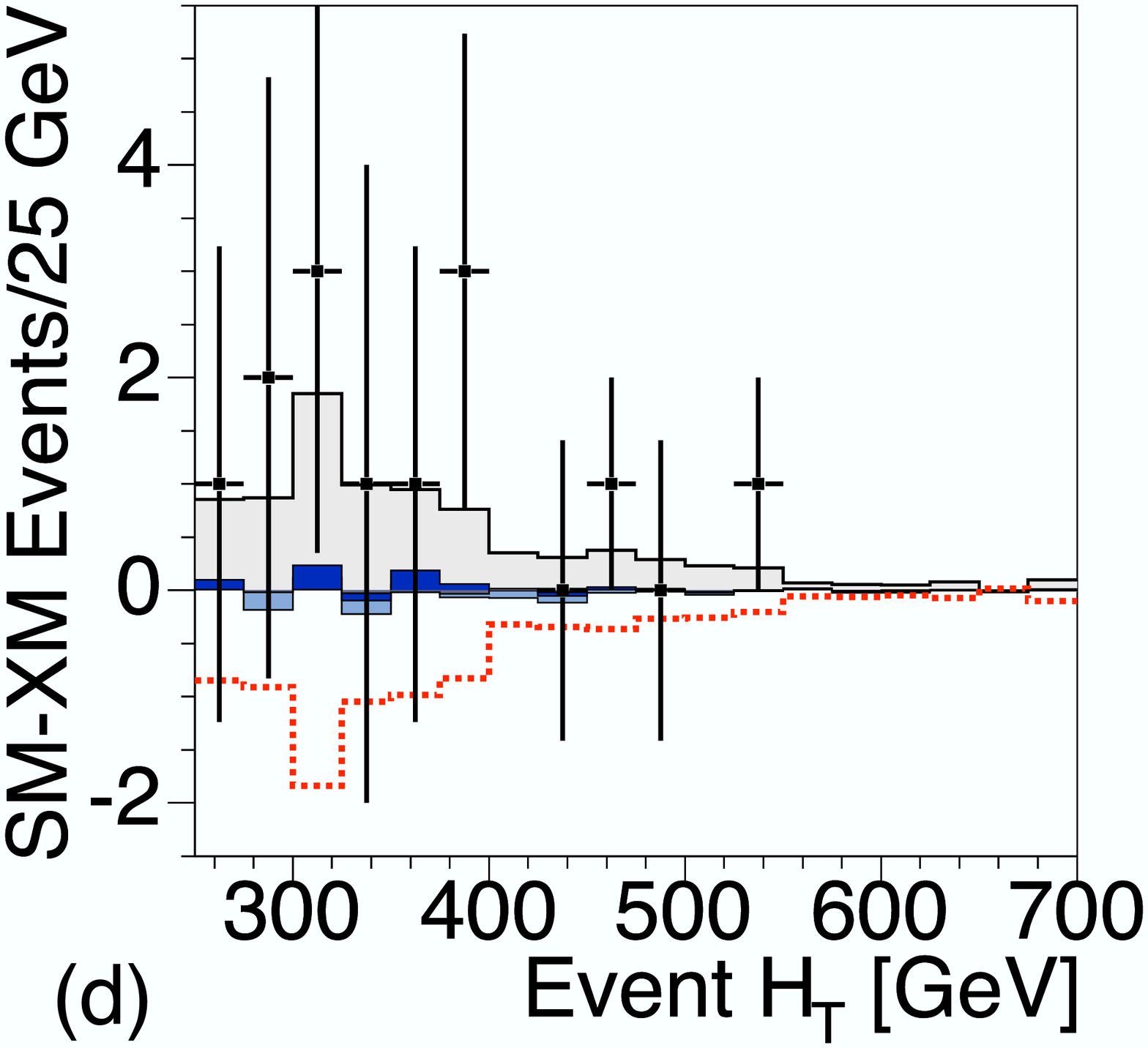}
\caption{Distribution of the SLT track $\pt$ for SM+XM tags (a), track $\pt$ for SM-XM tags (b), event $\Ht$ for SM+XM tags (c), and event $\Ht$ for SM-XM tags (d).  The expectation for a hypothetical XM top quark is shown in the dotted line.}
\label{fig:kinhist}
\end{figure}

Figure~\ref{fig:kinhist} shows the distribution of the event $\Ht$ and the SLT tag $\pt$.  Both the sum and difference of the events classified as SM and XM are shown.  The total $\ttbar$ contribution (SM+XM) from simulation is normalized to the data and divided between SLT contributions from direct semileptonic $b$ decay, cascade semileptonic decay, and other sources.  The expected distribution assuming a $-$4/3 charge XM top quark is shown as a dashed line in the SM-XM plots.  These figures demonstrate the preference of the asymmetry for the SM expectation as a function of the event kinematics.

In conclusion, we have presented the strongest exclusion of an exotic top quark with $-$4/3 charge to date (at 95\% C.L.), while observing strong evidence for the SM $+$2/3 electric charge of the top quark.  This measurement improves on both the expected and measured $p$-values reported in Ref.~\cite{JetQ}.  For purposes of comparison, we note that what is labeled as ``expected C.L.'' in Ref.~\cite{JetQ} corresponds to one minus the expectation value of our $p_{\rm XM}$ under the SM hypothesis.  This is the first time soft leptons tags have been used to accomplish such a measurement.

\begin{acknowledgments}
We thank the Fermilab staff and the technical staffs of the participating institutions for their vital contributions. This work was supported by the U.S. Department of Energy and National Science Foundation; the Italian Istituto Nazionale di Fisica Nucleare; the Ministry of Education, Culture, Sports, Science and Technology of Japan; the Natural Sciences and Engineering Research Council of Canada; the National Science Council of the Republic of China; the Swiss National Science Foundation; the A.P. Sloan Foundation; the Bundesministerium f\"ur Bildung und Forschung, Germany; the World Class University Program, the National Research Foundation of Korea; the Science and Technology Facilities Council and the Royal Society, UK; the Institut National de Physique Nucleaire et Physique des Particules/CNRS; the Russian Foundation for Basic Research; the Ministerio de Ciencia e Innovaci\'{o}n, and Programa Consolider-Ingenio 2010, Spain; the Slovak R\&D Agency; and the Academy of Finland.
\end{acknowledgments}

\end{document}

%% file: October2009_Authors.tex
\affiliation{Institute of Physics, Academia Sinica, Taipei, Taiwan 11529, Republic of China} 
\affiliation{Argonne National Laboratory, Argonne, Illinois 60439, USA} 
\affiliation{University of Athens, 157 71 Athens, Greece} 
\affiliation{Institut de Fisica d'Altes Energies, Universitat Autonoma de Barcelona, E-08193, Bellaterra (Barcelona), Spain} 
\affiliation{Baylor University, Waco, Texas 76798, USA} 
\affiliation{Istituto Nazionale di Fisica Nucleare Bologna, $^{dd}$University of Bologna, I-40127 Bologna, Italy} 
\affiliation{Brandeis University, Waltham, Massachusetts 02254, USA} 
\affiliation{University of California, Davis, Davis, California 95616, USA} 
\affiliation{University of California, Los Angeles, Los Angeles, California 90024, USA} 
\affiliation{University of California, San Diego, La Jolla, California 92093, USA} 
\affiliation{University of California, Santa Barbara, Santa Barbara, California 93106, USA} 
\affiliation{Instituto de Fisica de Cantabria, CSIC-University of Cantabria, 39005 Santander, Spain} 
\affiliation{Carnegie Mellon University, Pittsburgh, Pennsylvania 15213, USA} 
\affiliation{Enrico Fermi Institute, University of Chicago, Chicago, Illinois 60637, USA}
\affiliation{Comenius University, 842 48 Bratislava, Slovakia; Institute of Experimental Physics, 040 01 Kosice, Slovakia} 
\affiliation{Joint Institute for Nuclear Research, RU-141980 Dubna, Russia} 
\affiliation{Duke University, Durham, North Carolina 27708, USA} 
\affiliation{Fermi National Accelerator Laboratory, Batavia, Illinois 60510, USA} 
\affiliation{University of Florida, Gainesville, Florida 32611, USA} 
\affiliation{Laboratori Nazionali di Frascati, Istituto Nazionale di Fisica Nucleare, I-00044 Frascati, Italy} 
\affiliation{University of Geneva, CH-1211 Geneva 4, Switzerland} 
\affiliation{Glasgow University, Glasgow G12 8QQ, United Kingdom} 
\affiliation{Harvard University, Cambridge, Massachusetts 02138, USA} 
\affiliation{Division of High Energy Physics, Department of Physics, University of Helsinki and Helsinki Institute of Physics, FIN-00014, Helsinki, Finland} 
\affiliation{University of Illinois, Urbana, Illinois 61801, USA} 
\affiliation{The Johns Hopkins University, Baltimore, Maryland 21218, USA} 
\affiliation{Institut f\"{u}r Experimentelle Kernphysik, Karlsruhe Institute of Technology, D-76131 Karlsruhe, Germany} 
\affiliation{Center for High Energy Physics: Kyungpook National University, Daegu 702-701, Korea; Seoul National University, Seoul 151-742, Korea; Sungkyunkwan University, Suwon 440-746, Korea; Korea Institute of Science and Technology Information, Daejeon 305-806, Korea; Chonnam National University, Gwangju 500-757, Korea; Chonbuk National University, Jeonju 561-756, Korea} 
\affiliation{Ernest Orlando Lawrence Berkeley National Laboratory, Berkeley, California 94720, USA} 
\affiliation{University of Liverpool, Liverpool L69 7ZE, United Kingdom} 
\affiliation{University College London, London WC1E 6BT, United Kingdom} 
\affiliation{Centro de Investigaciones Energeticas Medioambientales y Tecnologicas, E-28040 Madrid, Spain} 
\affiliation{Massachusetts Institute of Technology, Cambridge, Massachusetts 02139, USA} 
\affiliation{Institute of Particle Physics: McGill University, Montr\'{e}al, Qu\'{e}bec, Canada H3A~2T8; Simon Fraser University, Burnaby, British Columbia, Canada V5A~1S6; University of Toronto, Toronto, Ontario, Canada M5S~1A7; and TRIUMF, Vancouver, British Columbia, Canada V6T~2A3} 
\affiliation{University of Michigan, Ann Arbor, Michigan 48109, USA} 
\affiliation{Michigan State University, East Lansing, Michigan 48824, USA}
\affiliation{Institution for Theoretical and Experimental Physics, ITEP, Moscow 117259, Russia} 
\affiliation{University of New Mexico, Albuquerque, New Mexico 87131, USA} 
\affiliation{Northwestern University, Evanston, Illinois 60208, USA} 
\affiliation{The Ohio State University, Columbus, Ohio 43210, USA} 
\affiliation{Okayama University, Okayama 700-8530, Japan} 
\affiliation{Osaka City University, Osaka 588, Japan} 
\affiliation{University of Oxford, Oxford OX1 3RH, United Kingdom} 
\affiliation{Istituto Nazionale di Fisica Nucleare, Sezione di Padova-Trento, $^{ee}$University of Padova, I-35131 Padova, Italy} 
\affiliation{LPNHE, Universite Pierre et Marie Curie/IN2P3-CNRS, UMR7585, Paris, F-75252 France} 
\affiliation{University of Pennsylvania, Philadelphia, Pennsylvania 19104, USA}
\affiliation{Istituto Nazionale di Fisica Nucleare Pisa, $^{ff}$University of Pisa, $^{gg}$University of Siena and $^{hh}$Scuola Normale Superiore, I-56127 Pisa, Italy} 
\affiliation{University of Pittsburgh, Pittsburgh, Pennsylvania 15260, USA} 
\affiliation{Purdue University, West Lafayette, Indiana 47907, USA} 
\affiliation{University of Rochester, Rochester, New York 14627, USA} 
\affiliation{The Rockefeller University, New York, New York 10021, USA} 
\affiliation{Istituto Nazionale di Fisica Nucleare, Sezione di Roma 1, $^{ii}$Sapienza Universit\`{a} di Roma, I-00185 Roma, Italy} 

\affiliation{Rutgers University, Piscataway, New Jersey 08855, USA} 
\affiliation{Texas A\&M University, College Station, Texas 77843, USA} 
\affiliation{Istituto Nazionale di Fisica Nucleare Trieste/Udine, I-34100 Trieste, $^{jj}$University of Trieste/Udine, I-33100 Udine, Italy} 
\affiliation{University of Tsukuba, Tsukuba, Ibaraki 305, Japan} 
\affiliation{Tufts University, Medford, Massachusetts 02155, USA} 
\affiliation{Waseda University, Tokyo 169, Japan} 
\affiliation{Wayne State University, Detroit, Michigan 48201, USA} 
\affiliation{University of Wisconsin, Madison, Wisconsin 53706, USA} 
\affiliation{Yale University, New Haven, Connecticut 06520, USA} 
\author{T.~Aaltonen}
\affiliation{Division of High Energy Physics, Department of Physics, University of Helsinki and Helsinki Institute of Physics, FIN-00014, Helsinki, Finland}
\author{J.~Adelman}
\affiliation{Enrico Fermi Institute, University of Chicago, Chicago, Illinois 60637, USA}
\author{B.~\'{A}lvarez~Gonz\'{a}lez$^w$}
\affiliation{Instituto de Fisica de Cantabria, CSIC-University of Cantabria, 39005 Santander, Spain}
\author{S.~Amerio$^{ee}$}
\affiliation{Istituto Nazionale di Fisica Nucleare, Sezione di Padova-Trento, $^{ee}$University of Padova, I-35131 Padova, Italy} 

\author{D.~Amidei}
\affiliation{University of Michigan, Ann Arbor, Michigan 48109, USA}
\author{A.~Anastassov}
\affiliation{Northwestern University, Evanston, Illinois 60208, USA}
\author{A.~Annovi}
\affiliation{Laboratori Nazionali di Frascati, Istituto Nazionale di Fisica Nucleare, I-00044 Frascati, Italy}
\author{J.~Antos}
\affiliation{Comenius University, 842 48 Bratislava, Slovakia; Institute of Experimental Physics, 040 01 Kosice, Slovakia}
\author{G.~Apollinari}
\affiliation{Fermi National Accelerator Laboratory, Batavia, Illinois 60510, USA}
\author{J.~Appel}
\affiliation{Fermi National Accelerator Laboratory, Batavia, Illinois 60510, USA}
% ...

\collaboration{CDF Collaboration\footnote{With visitors from $^a$University of Massachusetts Amherst, Amherst, Massachusetts 01003,
$^b$Universiteit Antwerpen, B-2610 Antwerp, Belgium, 
$^c$University of Bristol, Bristol BS8 1TL, United Kingdom,
$^d$Chinese Academy of Sciences, Beijing 100864, China, 
$^e$Istituto Nazionale di Fisica Nucleare, Sezione di Cagliari, 09042 Monserrato (Cagliari), Italy,
$^f$University of California Irvine, Irvine, CA  92697, 
$^g$University of California Santa Cruz, Santa Cruz, CA  95064, 
$^h$Cornell University, Ithaca, NY  14853, 
$^i$University of Cyprus, Nicosia CY-1678, Cyprus, 
$^j$University College Dublin, Dublin 4, Ireland,
$^k$University of Edinburgh, Edinburgh EH9 3JZ, United Kingdom, 
$^l$University of Fukui, Fukui City, Fukui Prefecture, Japan 910-0017,
$^m$Kinki University, Higashi-Osaka City, Japan 577-8502,
$^n$Universidad Iberoamericana, Mexico D.F., Mexico,
$^o$University of Iowa, Iowa City, IA  52242,
$^p$Kansas State University, Manhattan, KS 66506,
$^q$Queen Mary, University of London, London, E1 4NS, England,
$^r$University of Manchester, Manchester M13 9PL, England,
$^s$Muons, Inc., Batavia, IL 60510, 
$^t$Nagasaki Institute of Applied Science, Nagasaki, Japan, 
$^u$University of Notre Dame, Notre Dame, IN 46556,
$^v$Obninsk State University, Obninsk, Russia,
$^w$University de Oviedo, E-33007 Oviedo, Spain, 
$^x$Texas Tech University, Lubbock, TX  79609, 
$^y$IFIC(CSIC-Universitat de Valencia), 56071 Valencia, Spain,
$^z$Universidad Tecnica Federico Santa Maria, 110v Valparaiso, Chile,
$^{aa}$University of Virginia, Charlottesville, VA  22906,
$^{bb}$Bergische Universit\"at Wuppertal, 42097 Wuppertal, Germany,
$^{cc}$Yarmouk University, Irbid 211-63, Jordan,
$^{kk}$On leave from J.~Stefan Institute, Ljubljana, Slovenia, 
}}
\noaffiliation

%% file: topq.bbl
\begin{thebibliography}{9}
\bibitem{Eta}
   We use a transverse coordinate system, where $\pt=p\sin(\theta)$ and $\et=E\sin(\theta)$ are the momentum and energy measured transverse to the beamline, respectively.  We define the pseudorapidity variable, $\eta\equiv-\ln\tan{(\theta/2)}$.  We define $\displaystyle{\not}{\et}$ as the negative vector sum of the transverse energies in all the calorimeter cells, $-\Sigma E^i_{T}$.
\bibitem{JetReco}
    We reconstruct jets using a fixed-cone algorithm with a cone size of $\Delta R\equiv\sqrt{\Delta\eta^2+\Delta\phi^2}\le0.4$, where $\phi$ is the azimuthal angle.
\end{thebibliography}
